\documentclass[aps,prl,10pt,a4paper,twocolumn,showpacs,superscriptaddress,longbibliography]{revtex4-1}
\usepackage{CJK,graphicx,xcolor,longtable}
\usepackage{amssymb,amsmath,amsfonts}
\usepackage[colorlinks=true,citecolor=blue,urlcolor=blue]{hyperref}
\begin{document}
\begin{CJK}{UTF8}{bsmi}
\author{Yu-Chin Tzeng}\email{d102054002@mail.nchu.edu.tw}
\affiliation{Department of Physics, National Chung-Hsing University, Taichung 40227, Taiwan}
\affiliation{Department of Physics, National Taiwan University, Taipei 10617, Taiwan}
\author{Hiroaki Onishi}\email{onishi.hiroaki@jaea.go.jp}
\affiliation{Advanced Science Research Center, Japan Atomic Energy Agency, Tokai, Ibaraki 319-1195, Japan}
\author{Tsuyoshi Okubo} 
\affiliation{Department of Physics, University of Tokyo, Tokyo 113-0033, Japan} 
\author{Ying-Jer Kao}\email{yjkao@phys.ntu.edu.tw}
\affiliation{Department of Physics, National Taiwan University, Taipei 10617, Taiwan}
\affiliation{National Center for Theoretical Sciences, National Tsing Hua University, Hsinchu 300, Taiwan}

\title{Quantum phase transitions driven by rhombic-type single-ion anisotropy\\ in the \textit{S}=1 Haldane chain}
\begin{abstract}
The spin-1 Haldane chain is an example of the symmetry-protected-topological (SPT) phase in one dimension. Experimental realization of the spin chain materials usually involves both the uniaxial-type, $D(S^z)^2$, and the rhombic-type, $E[(S^x)^2-(S^y)^2]$, single-ion anisotropies. Here, we provide a precise ground-state phase diagram for spin-1 Haldane chain with these single-ion anisotropies. Using quantum numbers, we find that the $\mathbb{Z}_2$ symmetry breaking phase can be characterized by double degeneracy in the entanglement spectrum. Topological quantum phase transitions take place on particular paths in the phase diagram, from the Haldane phase to the Large-$E_x$, Large-$E_y$, or Large-$D$ phases. The topological critical points are determined by the level spectroscopy method with a newly developed parity technique in the density matrix renormalization group [{\color{blue}Phys. Rev. B \textbf{86}, 024403 (2012)}], and the Haldane-Large-$D$ critical point is obtained with an unprecedented precision, $(D/J)_c$=0.9684713(1). Close to this critical point, a small rhombic single-ion anisotropy $|E|/J\ll1$ can destroy the Haldane phase and bring the system into a $y$-N\'eel phase. We propose that the compound [Ni(HF$_2$)(3-Clpy)$_4$]BF$_4$ is a candidate system to search for the $y$-N\'eel phase.
\end{abstract}
\date \today
\maketitle
\end{CJK}

\textit{Introduction.}
Quantum magnetism of integer-spin chains has been attracting attention for decades. It was stimulated by the Haldane conjecture~\cite{Haldane:PRL1983} that the lowest excitation in the antiferromagnetic Heisenberg model are gapped if and only if the spin~$S$ is an integer. Experimental evidences for the Haldane gap were discovered in several $S$=1 quasi-one dimensional (Q1D) materials, \textit{e.g.}, $\mathrm{CsNiCl}_3$~\cite{CsNiCl3:1986,CsNiCl3:2002}, Y$_2$BaNiO$_5$~\cite{Y2BaNiO5,YBNO:2013,YBNO:2015}, Ni(C$_2$H$_8$N$_2$)$_2$NO$_2$(ClO$_4$) (NENP)~\cite{NENP,NENP:Eterm}, and [Ni(C$_2$H$_8$N$_2$)$_2$NO$_2$]BF$_4$ (NENB)~\cite{NENB}. 
Due to the crystal field and the spin-orbit coupling, the microscopic effective Hamiltonian for the Q1D spin chains involves the single-ion anisotropies,
\begin{equation}\label{eq:Ham}
  H=J\sum_{i=1}^L\vec S_i \cdot \vec S_{i+1}+D\sum_{i=1}^L(S_i^z)^2+E\sum_{i=1}^L[(S_i^x)^2-(S_i^y)^2],
\end{equation}
where $J$\textgreater0 is the strength of the Heisenberg exchange interaction, as well as $D$ and $E$ are the parameters of the uniaxial and rhombic single-ion anisotropies, respectively. The Haldane gap is robust against small anisotropies, and it extends to a region so-called Haldane phase. In the absence of a local order, the Haldane phase falls beyond the paradigm of Landau's theory of phase transitions. From a topological point of view, the Haldane phase is classified as the symmetry-protected-topological (SPT) phase~\cite{SPT} for odd-$S$, while the Haldane phase is adiabatically connected with a topological trivial phase for even-$S$~\cite{Pollmann2012,Tonegawa2011,Okamoto2016,Tzeng2012,Kjall2013}. Interesting properties such as the Valence-Bond-Solid (VBS) description~\cite{AKLT1987}, hidden $\mathbb{Z}_2\times\mathbb{Z}_2$ symmetry breaking, non-local string order, fractionalized gapless edge modes, and the degenerate entanglement spectra are used to characterize the SPT phase.
On the other hand, the entanglement spectrum is not required to be degenerate for both the topological trivial phase and the symmetry breaking phase.


\begin{figure}[b]
\centering
\includegraphics[width=3.4in]{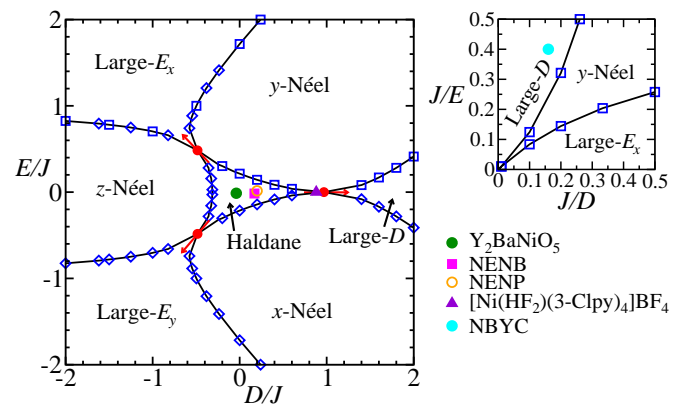} 
\caption{Quantum phase diagram for $S$=1 Haldane chain with both uniaxial and rhombic single-ion anisotropies.
Topological quantum phase transitions occur through particular (red arrows) routes. The Haldane-Large-$D$ critical point $(D/J)_c$=0.9684713(1) is determined by the LS+DMRG method. A small rhombic anisotropy $|E|/J\ll1$ at this point $(D/J)_c$ induces a transverse antiferromagnetic order.
}\label{Fig:phase}
\end{figure}

Prior theoretical and numerical investigations focus on the effects of the uniaxcial anisotropy ($D$-term)~\cite{WChen2003,Boschi2003,Boschi2006,Tzeng2008a,Tzeng2008b,Pollmann2010,Cui2013,Braganca2014,Yousef2015,Fuji2016}.
The effect of rhombic anisotropy ($E$-term) lacks a complete theoretical understanding~\cite{Tonegawa1995,Rudowicz2014}; however, materials with large $D/J$ and $E/J$ are discovered, \textit{e.g.}, the $S$=1 Q1D chains, Sr$_3$NiPtO$_6$~\cite{SNPO,SNPO:DFT}, Ni(C$_2$H$_8$N$_2$)$_2$Ni(CN)$_4$ (NENC)~\cite{NENC}, [Ni(HF$_2$)(3-Clpy)$_4$]BF$_4$ (py=pyridine)~\cite{D088,1409.5971}, and Ni(C$_{10}$H$_8$N$_2$)$_2$Ni(CN)$_4\cdot$H$_2$O (NBYC)~\cite{NBYC:2000,NBYC:2004,strongE2004b}.

In this Rapid~Communication, we fill up the vacancy in the survey of the phase diagram regarding the $E$-term. By means of the DMRG~\cite{White1992}, within the periodic boundary conditions (PBC), the ground-state phase diagram of the $S$=1 Hamiltonian Eq.~(\ref{eq:Ham}) is shown in the Fig.~\ref{Fig:phase}, and some of Q1D spin-1 materials are listed in Tab.~\ref{tab:list}. By the permutations of spin operators, the phase diagram shows a ``rotational'' symmetry in the rescaled $D$-$\sqrt{3}E$ parameter space~\cite{supp}. The Hamiltonian does not conserve the magnetization $M$=$\sum_iS_i^z$ because of the $E$-term. Instead of the magnetization, the parity of $M$ is conserved. $m$=$M\bmod2$=\ 0~or~1, is a good quantum number since the $E$-term raises or lowers the magnetization by~2. The spatial inversion $p$=$\pm1$ and time reversal $t$=$\pm1$ are also good quantum numbers. We label the energy eigenstates and the entanglement states by these quantum numbers $(m,p,t)$. The number of states kept $K$ is up to 2000 in this study.

\begin{table}[t]
\caption{\label{tab:list}
The values of zero-field-splitting parameters for some spin-1 Q1D materials.}
\begin{ruledtabular}
\begin{tabular}{ l c c c c }
Compounds & $D/J$ & $E/J$ & Phase & Ref.\\
\colrule
Y$_2$BaNiO$_5$ & $-0.039$ & $-0.0127$ & Haldane & [\onlinecite{Y2BaNiO5}] \\
NENB & 0.17 & $-0.016$ & Haldane & [\onlinecite{NENB}] \\
NENP & 0.2 & 0.01-0.02 & Haldane & [\onlinecite{NENP:Eterm}] \\
$[\mathrm{Ni}(\mathrm{HF}_2)\mbox{(3-Clpy)}_4]\mathrm{BF}_4$ & 0.88 &   &   & [\onlinecite{D088},\onlinecite{1409.5971}] \\
NBYC\footnote{Susceptibility.} & 6.25 & 2.5 & Large-$D$ & [\onlinecite{NBYC:2000}] \\
NBYC\footnote{An additional bilinear-biquadratic term is considered.} & 7.49 & 4.26 & Large-$D$ & [\onlinecite{NBYC:2004}] \\
NENC & 7.5 & 0.83 & Large-$D$ & [\onlinecite{NENC}] \\
Sr$_3$NiPtO$_6$ & 8.8 & 0 & Large-$D$ & [\onlinecite{SNPO},\onlinecite{SNPO:DFT}] \\
\end{tabular}
\end{ruledtabular}
\end{table}

\textit{Energy and Entanglement Spectrum.} 
The Haldane phase surrounded by the other phases is a SPT phase~\cite{SPT} protected by the dihedral group, the time reversal, and the space-inversion symmetries~\cite{Pollmann2010,Pollmann2012}.
The ground state can be described by the VBS picture~\cite{AKLT1987}:
Each spin-1 in the chain is regarded as triplet states of two spin-1/2, and the neighboring spin-1/2 of different spin-1 form a valence bond, the singlet state. From the VBS picture, two consequences are inferred.
First, because each singlet contributes odd quantum numbers for both spatial inversion and time reversal, a closed chain of even number of singlets has quantum numbers $(m,p,t)=(0,1,1)$. Therefore, we compute the ground state energy $E_g=E_0(0,1,1;\mathrm{PBC})$ in this sector. Second, with the Haldane gap in the bulk, an open chain has free unpaired spin-1/2 states at the edges. For the PBC, a closed chain in our case, the edge states can be artificially created by the partial trace of one part of the bipartition. Explicitly, the chain is divided into two subsystems A and B with equal sizes, and the reduced density matrix $\rho_A$=$\mathrm{Tr}_B|\psi_0\rangle\langle\psi_0|$ is computed in the DMRG, where $|\psi_0\rangle$ is the ground state. The entanglement spectrum is defined by $\xi_i$=$-\ln\omega_i$, where $\omega_i$ is the $i$-th largest eigenvalue of $\rho_A$. The edge states reflect on that the reduced state can be decomposed into the product $\rho_A\approx(\frac{1}{2}\openone_{2\times2})\otimes\rho_0\otimes(\frac{1}{2}\openone_{2\times2})$~\cite{Tzeng2016,PRX2016}, where $\openone_{2\times2}$ are the two-by-two identity matrices of the edges, the boundary between A~and~B, and $\rho_0$ is a pure-state bulk-part matrix of the subsystem~A. This fact ensures a four-fold degenerate entanglement spectrum in the Haldane phase.
The four-fold degeneracy can be seen as a simple illustration of the bulk-edge correspondence in the entanglement spectrum~\cite{bulkedge,edgeES,ES}.

\begin{figure}[t]
\centering
\includegraphics[width=3.4in]{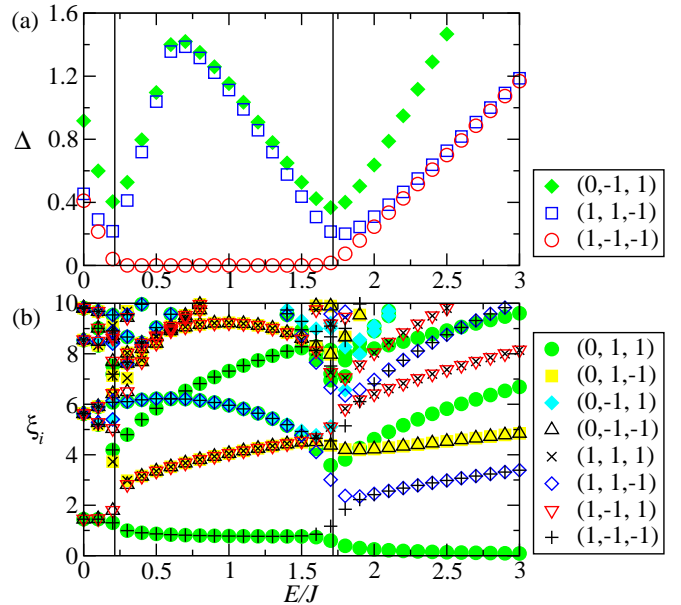} 
\caption{
(a) The excitation gap $\Delta$=$E_0(m,p,t;\mathrm{PBC})-E_g$, where $E_g$=$E_0(0,1,1;\mathrm{PBC})$ is the ground state energy. The ground state in the $y$-N\'eel phase has double degeneracy.
(b) The entanglement spectra of at least four-fold, double, and non-degeneracy characterize the Haldane, the $y$-N\'eel, and the Large-$E_x$ phases, respectively.
The vertical lines indicate two critical points $(E/J)_c\simeq0.214$ and $1.717$, respectively.
DMRG data for $D/J$=0 and $L$=80 are presented. $K$=1000 states are kept.
}\label{Fig:gap}
\end{figure}

\begin{figure}[t]
\centering
\includegraphics[width=3.4in]{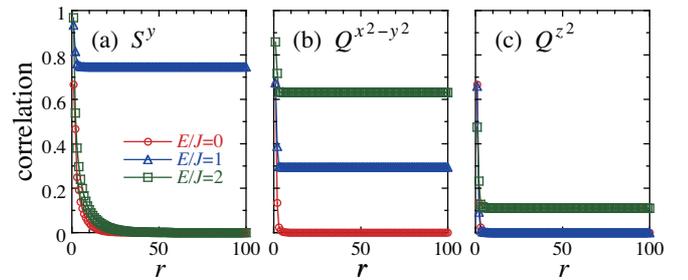} 
\caption{Spin and quadrupole correlation functions at typical values of $E/J$ for $D/J$=0 and $L$=200. (a) $(-1)^r\langle{S}_0^yS_r^y\rangle$, (b) $\langle{Q}_0^{x^2-y^2}Q_r^{x^2-y^2}\rangle$, and (c) $\langle{Q}_0^{z^2}Q_r^{z^2}\rangle$. Note that $r$=100 for the most distant sites in the periodic chain of $L$=200.}
\label{Fig:corr}
\end{figure}

The lowest energy excitations and the low-lying entanglement spectra are shown in the Fig.~\ref{Fig:gap} for fixed $D/J$=0. The Haldane gap is estimated as $0.41J$ for $D$=$E$=0, which agrees with recent numerical results~\cite{Ejima2015}. Double degeneracy in the ground-state energy and the entanglement spectrum are found in the region between the Haldane phase and the Large-$E_x$ phase. Both of the double degeneracies come from the nature of a $\mathbb{Z}_2$ spontaneous symmetry breaking phase, with breaking the parity of magnetization, space inversion, and time-reversal symmetries, simultaneously. The spin-spin correlation shows the phase also breaks translational symmetry, as we will see in Fig.~\ref{Fig:corr}, therefore we refer to this phase as the $y$-N\'eel phase.

The degeneracy structure of entanglement spectrum has been proposed to distinguish different many-body quantum phases recently~\cite{ES,Pollmann2010,Li2013,Singh2015,Saadatmand2016}.
The reason of the double degeneracy in the $y$-N\'eel phase, Fig.~\ref{Fig:gap}(b), is that, the degenerate ground state is selected as an eigenstate of the symmetry operators by the quantum numbers $(m,p,t)$ in the DMRG~\cite{Tzeng2012}. Such an enforced symmetrized state is similar to a Greenberger-Horne-Zeilinger (GHZ) state, and the artificial double degenerate spectrum is generated. For example, the state $\frac{1}{\sqrt{2}}(\mid\nearrow\hspace{-1.5mm}\swarrow\hspace{-1.5mm}\nearrow\hspace{-1.5mm}\swarrow\rangle\pm\hspace{-1.4mm}\mid\swarrow\hspace{-1.5mm}\nearrow\hspace{-1.5mm}\swarrow\hspace{-1.5mm}\nearrow\rangle)$ has the inversion parity quantum number $p$=$\pm1$, and it is an entangled state. However, each symmetry breaking state, $\mid\nearrow\hspace{-1.5mm}\swarrow\hspace{-1.5mm}\nearrow\hspace{-1.5mm}\swarrow\rangle$ or $\mid\swarrow\hspace{-1.5mm}\nearrow\hspace{-1.5mm}\swarrow\hspace{-1.5mm}\nearrow\rangle$, is not an eigenstate of the symmetry operator, and it is not entangled, either. Note that the double degeneracy appears in the entire spectrum, and is also found in the $z$-N\'eel and $x$-N\'eel phases.
Thus, the degeneracy structure of the entanglement spectrum identifies the SPT phase (four-fold), the $\mathbb{Z}_2$ symmetry breaking phase (double), and the topological trivial phase (single).

In the $z$-N\'eel phase, the parity of magnetization $m$ is conserved, therefore the other parity quantum numbers $(p,t)$ are essential for observing the degenerate spectrum. In contrast, in the $y$-N\'eel (or $x$-N\'eel) phase, the double degeneracy can be observed when only using the parity of magnetization. Therefore the $E$-term serves an ideal model Hamiltonian to observe the parity degeneracy in the spontaneous symmetry breaking phase.
From the technical point of view, quantum numbers are usually used in the DMRG for preventing the mixture of different subspaces as well as stabilizing and accelerating the computations. Because the magnetization is the most often used quantum number in the DMRG, programming with the parity of magnetization should be easier than the parity of inversion or time-reversal.

\textit{Correlation Functions.--}
The microscopic spin states of novel quantum phases induced by single-ion anisotropies can be clarified by measuring spin correlation functions, $\langle{S}_{0}^{\alpha}{S}_{r}^{\alpha}\rangle$, and quadrupole correlation functions, $\langle{Q}_{0}^{\gamma}{Q}_{r}^{\gamma}\rangle$, where $Q_{i}^{x^{2}-y^{2}}=(S_{i}^{x})^{2}-(S_{i}^{y})^{2}$ and $Q_{i}^{z^{2}}=\frac{1}{\sqrt{3}}[3(S_{i}^{z})^{2}-2]$ are relevant quadrupole operators. In Fig.~\ref{Fig:corr}, we show typical behavior with increasing $E/J$ for fixed $D/J$=0. In the Haldane phase for small anisotropies, spin and quadrupole correlations are short-ranged, as shown for $E/J$=0.
At intermediate $E/J$, robust antiferromagnetic correlations of the spin $y$-component occur, as shown for $E/J$=1. Spin correlations of the $x$- and $z$-components are short-ranged (not shown). Note here that the local spin state is forced to be the lowest-energy eigenstate of $E[(S_i^x)^2-(S_i^y)^2]$, given by $\vert{S}_i^x$=0$\rangle$, where the local spin fluctuates in the $yz$-plane. Such fluctuating spins align antiferromagnetically, while they preferably point to the $y$-direction due to the $E$-term. Thus the $y$-N\'eel phase is identified.

At large $E/J$, the ground state turns to be the product of $\vert{S}_i^x$=0$\rangle$, and the N\'eel structure vanishes, as shown for $E/J$=2. This phase is referred to as the Large-$E_x$ phase.
Because the negative $E/J$ is equivalent to exchanging the $x$-axis and $y$-axis, $E[(S_i^y)^2-(S_i^x)^2]$, we refer the phase as the Large-$E_y$ phase for the product of $\vert{S}_i^y$=0$\rangle$ at large negative $E/J$, and only positive $E/J$ is discussed.
A distinct feature for the Large-$E_x$ phase is that $\vert{S}_i^x$=0$\rangle$ has quadrupole moments, $\langle{Q}_i^{x^2-y^2}\rangle=-1$ and $\langle{Q}_i^{z^2}\rangle=\frac{1}{\sqrt{3}}$, so that finite quadrupole correlations come out. The $Q^{x^2-y^2}$ correlation develops in the $y$-N\'eel and Large-$E_x$ phases, while the $Q^{z^2}$ correlation grows after entering the Large-$E_x$ phase.
We should note that the $E$- and $D$-terms have the same forms as $Q^{x^2-y^2}$ and $Q^{z^2}$, respectively, indicating emergent quadrupole degrees of freedom. We expect that the competition of quadrupole states would drive the system into quadrupole phases, the so-called spin nematic phases, but N\'eel phases are observed in the present case at zero magnetic field. The search for possible quadrupole phases in magnetic field would be an interesting future problem, since those in a spin-1/2 frustrated chain in high magnetic field have been actively discussed~\cite{Chubukov1991,Hikihara2008,Onishi2015a,Onishi2015b,Masuda2011,Mourigal2012}.

\textit{Level Spectroscopy.}
The critical points are determined by the finite size scaling of the entanglement entropy~\cite{Amico:RMP,Vidal2003,Korepin2004,Renyi,supp} and the level spectroscopy (LS) method. All the transitions belong to the Ising universality class with the central charge $c$=$\frac{1}{2}$, except three Gaussian points with $c$=1 labelled by the red points in Fig.~\ref{Fig:phase}. Topological quantum phase transitions occur at these Gaussian points, from the Haldane phase to the Large-$D$, Large-$E_x$, or Large-$E_y$ phases. The topological quantum phase transition from the Haldane phase to the Large-$D$ phase is known as an example of the third-order Gaussian transition~\cite{Tzeng2008b}, therefore this critical point is more difficult to be precisely determined than the conventional second-order transitions. Severial methods for the determination of this critial point were investigated, including the LS plus exact diagonalization (LS+ED)~\cite{WChen2003}, fidelity susceptibility~\cite{Tzeng2008a,Tzeng2008b}, quantum monte calro (QMC)~\cite{QMC}, von Neumann entropy~\cite{accurate}, and the quantum renormalization group.~\cite{QRG} Here we use the parity DMRG~\cite{Tzeng2012} to perform the LS+DMRG method.

\begin{figure}[t]
\centering
\includegraphics[width=3.2in]{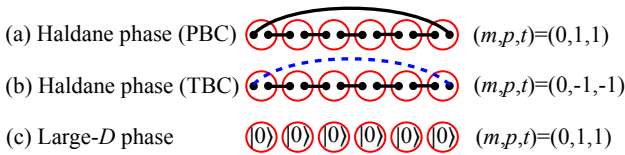}
\caption{(a) For a closed chain with even number of singlets, the quantum numbers for the Haldane phase are $(m,p,t)=(0,1,1)$. (b) Within TBC, the number of singlets become odd, and the quantum numbers change as $(m,p,t)=(0,-1,-1)$. (c) For large single-ion ansiotropy, the quantum numbers are $(m,p,t)=(0,1,1)$.}\label{Fig:lspic}
\end{figure}

The LS method~\cite{Nomura1995,Kitazawa1997,Kitazawa1997a,Kitazawa1997b,Nomura1998,howto} is based on the effective field theory of the sine-Gordon model and the $c$=1 conformal field theory. The critical point can be probed by the energy level crossing within the twisted boundary conditions (TBC), $S_{L+1}^x\to-S_1^x$, $S_{L+1}^y\to-S_1^y$, $S_{L+1}^z\to{S}_1^z$.
The LS method can be roughly described by the VBS picture~\cite{howto}, as shown in Fig.~\ref{Fig:lspic}.
For the TBC chain with even length $L$, there are odd number of singlet bonds and one triplet bond in the Haldane phase. Each singlet contributes the inversion parity quantum number $p_i=-1$, and the triplet bond contributes $p_L=1$. Thus, the quantum number for the system becomes odd, $p=\prod_{i=1}^Lp_i=-1$. On the other hand, the inversion parity quantum number for the Large-$D$ phase is always even, $p=1$. Therefore, the Haldane phase and the Large-$D$ phase are characterized by the energy $E_0(0,-1,-1;\mathrm{TBC})$ and $E_0(0,1,1;\mathrm{TBC})$, respectively.

\begin{figure}[t]
\centering
\includegraphics[width=3.4in]{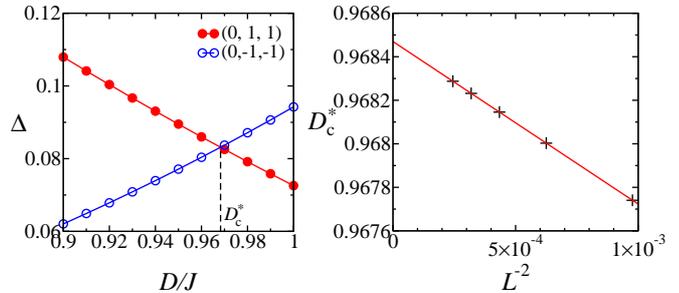} 
\caption{(Left) Energy level crossing with different quantum numbers occurs at $D_c^*$ for $L$=$64$, $E/J$=0, within TBC. $\Delta$=$E_0(m,p,t;\mathrm{TBC})-E_g$ and $E_g$=$E_0(0,1,1;\mathrm{PBC})$ is the ground state energy within PBC.
(Right) Extrapolation of the critical point is performed by linear fitting. $(D/J)_c$=$0.9684713(1)$ is obtained.
$K$=2000 states are kept.}\label{Fig:LS}
\end{figure}

We show the energy level crossing $E_0(m,p,t;\mathrm{TBC})$ with different quantum numbers $(m,p,t)$=$(0,1,1)$ and $(0,-1,-1)$ in Fig.~\ref{Fig:LS}. The location of the crossing point is labelled by $D/J$=$D_c^*$, and critical point is obtained by the extrapolation to the thermodynamic limits. It is known that the scaling formula is a polynomial function in $L^{-2}$~\cite{Nomura1995,Kitazawa1997,Kitazawa1997a,Kitazawa1997b,Nomura1998,howto}. This importantly makes the convergence fast, because the subleading term $L^{-4}$ is much smaller than the leading term. Our numerical data~\cite{supp}, for $L=32$, $40$, $48$, $56$, and $64$, show that a linear fitting is good enough for the extrapolation. We obtain $(D/J)_c=0.96847133(2)$ with linear fitting and $(D/J)_c=0.96847141(2)$ with the subleading term $L^{-4}$. Therefore it would be safe to conclude $(D/J)_c=0.9684713(1)$ with the systematic error about $10^{-7}$. Although our LS+DMRG only have sizes $L\leq64$, combining the DMRG technique proposed by~\citet{accurate} for large systems with level spectroscopy should further improve the precision of the value $(D/J)_c$. Other combinations such as LS+QMC~\cite{Suwa2015} are also possible.

Finally, we briefly discuss the effect of $E$-term at these (red points in Fig.~\ref{Fig:phase}) topological critical points.
Basically, the effect of $E$-term can be understood from the scaling dimension of the $E$-term at the Haldane-Large-$D$ critical point $(D/J)_c$ in the renormalization group flow analysis. If the $E$-term is relevant at the critical point, the critical point disappears by introducing an infinitesimal $E$-term as observed in the phase diagram in $D$-$E$ plane, Fig.~\ref{Fig:phase}.
The Haldane-Large-$D$ critical point $(D/J)_c$ is characterized by the central charge $c$=1 free boson conformal field theory~\cite{Boschi2003}. Note that the $E$-term can be transformed as
\begin{equation}
 E\sum_{i=1}^L[(S_i^x)^2-(S_i^y)^2]=\frac{E}{2}\sum_{i=1}^L[(S_i^+)^2+(S_i^-)^2].
\end{equation}
Thus, if the scaling dimension of $(S_i^+)^2$ at the critical fixed point is less than the dimension $1+1=2$, the $E$-term is relevant and an infinitesimal $E/J$ flows away from the critical point. The renormalization flow may eventually goes to $x$- or $y$-N\'{e}el phases depending on the sign of $E/J$.
Actually, a recent DMRG calculation has estimated the scaling dimension $\Delta_s$ corresponding to this operator as $\Delta_s=0.750\pm0.002$~\cite{Boschi2003}. Because the scaling dimension clearly satisfies the relation $\Delta_s$\textless2, we can conclude that the effect of $E$-term is relevant at the Haldane-Large-$D$ critical fixed point $(D/J)_c$. Thus we expect that by introducing infinitesimal $E/J$, the critical point between the Haldane and the Large-$D$ phases disappears because the relevant $E$-term increases along renormalization and it flows away from the critical point. By considering the symmetries of permutation of axis~\cite{supp}, the present discussions are also applicable for the critical points between the Haldane phase and Large-$E_x$ or Large-$E_y$ phases.

\textit{Discussions.}
We have investigated and provided a precise quantum phase diagram for the $S$=1 Haldane chain with both uniaxial and rhombic single-ion anisotropies, Eq.~(\ref{eq:Ham}). By the parity DMRG~\cite{Tzeng2012} within PBC, we show that, for the first time, the symmetry breaking phase has double degeneracy in the entire entanglement spectrum. This generalize the perspective that the degeneracy structure of entanglement spectrum tells different quantum phases, from the SPT phases to the symmetry breaking phases. The Haldane-Large-$D$ critical point is determined by the LS+DMRG method with an unprecedented accurate value $(D/J)_c=0.9684713(1)$. The presented power of the LS+DMRG method supports the reliability of finding the SPT intermediate-$D$ phase in $S$=2 XXZ chain~\cite{Tonegawa2011,Okamoto2016,Tzeng2012}. From the phase diagram, we point out that a small rhombic anisotropy induces a transverse antiferromagnetic long range order when $D/J$ close to this $(D/J)_c$. This suggests that [Ni(HF$_2$)(3-Clpy)$_4$]BF$_4$, with $D/J\simeq0.88$~\cite{D088}, is either a possible candidate system to search for the $y$-N\'eel phase, or a candidate for observing the quantum phase transition driven by the rhombic-type single-ion anisotropy.

In the end of this Rapid Communication, we argue the spin-1 chain can be made by arranging the single-molecule-magnets (SMM), \textit{e.g.}, CoH, the $S$=1 SMM~\cite{CoH}. We mention that recent experiments on a small cluster of SMMs have been taken into account the weak interactions between SMMs for $L$=2~\cite{L2} and $L$=4~\cite{L4}. On the other hand, atomic engineering has been able to tune the magnetic anisotropy~\cite{Hunds} and tune the spin state by absorbing hydrogen~\cite{CoH,srep2013}. The spin-spin interaction coming from the superexchange mechanism~\cite{CuN,DFT,Otto2009,Loth2012} and the Ruderman-Kittel-Kasuya-Yosida (RKKY) interaction~\cite{RKKY} have been observed. Cold rubidium atoms have recetly been proposed to simulate a spin-1 chain with uniaxcial-type single-ion anisotropy~\cite{coldatom}. In principle, an artificial spin chain with both uniaxial and rhombic single-ion anisotropies can be created.

\textit{Acknowledgements.} 
Y.C.T. and Y.J.K. are grateful to MOST in Taiwan via No. MOST105-2112-M-002-023-MY3. H.O. acknowledges the support of JSPS KAKENHI Grant Number JP16K05494. We are grateful to the National Center for High-performance Computing for computer time and facilities. Computations were also done on the supercomputers at the Japan Atomic Energy Agency and the Institute for Solid State Physics, the University of Tokyo. We are grateful to Hui Zhai, Kwai-Kong Ng, Markus Ternes and Mark Meisel for many useful discussions. T.O. was supported by JSPS KAKENHI Grant Number JP15K17701 and by MEXT of Japan as a social and scientific priority issue (Creation of new functional devices and high-performance materials to support next-generation industries; CDMSI) to be tackled by using post-K computer.
%

\clearpage
\appendix
\begin{widetext}
\section{Supplementary Materials: Quantum phase transitions driven by rhombic-type single-ion anisotropy in the \textit{S}=1 Haldane chain}
\begin{center}
Yu-Chin Tzeng$^{1,2}$, Hiroaki Onishi$^{3}$, Tsuyoshi Okubo$^{4}$, Ying-Jer Kao$^{2,5}$\\
\quad\\
$^1$\textit{Department of Physics, National Chung-Hsing University, Taichung 40227, Taiwan}\\
$^2$\textit{Department of Physics, National Taiwan University, Taipei 10617, Taiwan}\\
$^3$\textit{Advanced Science Research Center, Japan Atomic Energy Agency, Tokai, Ibaraki 319-1195, Japan}\\
$^4$\textit{Department of Physics, University of Tokyo, Tokyo 113-0033, Japan}\\
$^5$\textit{National Center of Theoretical Sciences, National Tsing Hua University, Hsinchu 300, Taiwan}
\end{center}
\end{widetext}

\begin{figure}[t]
\centering
\includegraphics[width=3.4in]{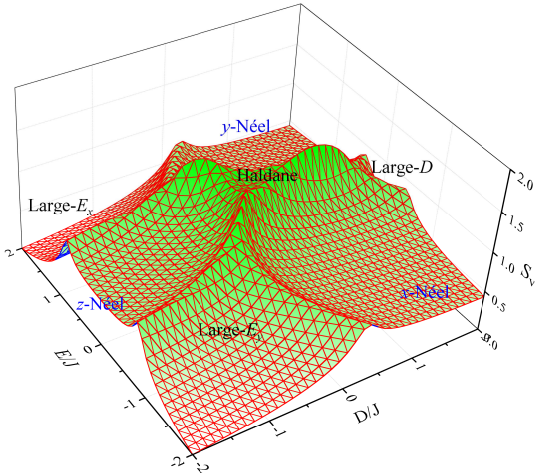} 
\caption{The von Neumann entropy in the parameter space for $L=40$. The subsystem size is $L_A=L/2$. The entropy diverges at the critical points in the thermodynamic limits, and remains a finite peaks in the finite size.
}\label{Fig:von}
\end{figure}

\subsection*{Entanglement entropy}
\begin{figure}[t]
\centering
\includegraphics[width=3.4in]{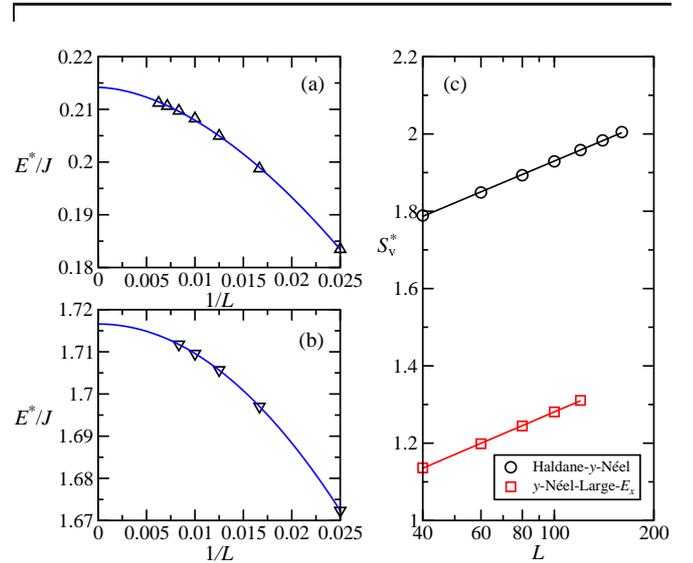} 
\caption{Finite size scaling of the von Neumann entropy for fixed $D=0$. (a) The Haldane-$y$-N\'eel critical point is about $(E/J)_c\simeq0.214$.
(b) The $y$-N\'eel-Large-$E_x$ critical point is about $(E/J)_c\simeq1.717$.
(c) The central charges are obtained by $c\simeq0.469$ and $c\simeq0.477$ for Haldane-$y$-N\'eel and $y$-N\'eel-Large-$E_x$ transitions, respectively.
}\label{Fig:fit}
\end{figure}

The entanglement in many-body systems has been developing as tools and perspective to the quantum critical phenomena~\cite{Amico:RMP}. The entanglement entropy is designed for quantifying the entanglement between A and B. The R\'enyi entropies $S_\alpha=\frac{1}{1-\alpha}\ln(\sum_i\omega_i^\alpha)$ and the von Neumann entropy $S_\mathrm{v}=-\sum_i\omega_i\ln\omega_i$ are two common used entropies, and it is convenient to be calculated in the DMRG. It was shown that these entropies diverge logarithmically with the subsystem size $L_A$ at the gapless quantum critical points, $S_\mathrm{v}\propto\frac{c}{3}\log L_A$, where $c$ is the central charge in the conformal field theory~\cite{Vidal2003,Korepin2004,Renyi}. In principle, the R\'enyi entropies and the lowest entanglement spectrum $\xi_0$ are also able to obtain the same critical points~\cite{Tzeng2016}; however, here we only perform the finite size scaling on the von Neumann entropy with the subsystem size $L_A=L/2$. The von Neumann entropy for $L=40$ in the $D$-$E$ parameter space is shown in Fig.~\ref{Fig:von}, and it is clear that the von Neumann entropy has a peak at the critical point. We label the value and the location of the peak as $S_\mathrm{v}^*$ and $E^*/J$ (or $D^*/J$ for fixed $E/J$) for fixed $D/J$, respectively. As shown in Fig.~\ref{Fig:fit}, after the finite size scaling, the critical points for fixed $D/J=0$ are obtained by $(E/J)_c\simeq0.214$ and $(E/J)_c\simeq1.717$. The corresponding central chagre are $c\simeq0.469$ and $c\simeq0.477$, respectively, and this indicates both the transitions belong to the Ising universality class. We obtain the central charge $c\approx\frac{1}{2}$ for all the critical points except three points, which are labeled by the red points in the phase diagram Fig.~\ref{Fig:phase}(a). It is known that the transition from the Haldane phase to the Large-$D$ phase is a Gaussian transition with the central charge $c=1$~\cite{Boschi2003,Boschi2006}. It is suitable to locate this critical point by the level spectroscopy method.

\begin{table}[t]
\caption{\label{tab:LS}
Numerical data for the location of the level crossing $D_c^*$ with different finite sizes. $K$ is the number of state kept in the DMRG.}
\begin{ruledtabular}
\begin{tabular}{ l l c c }
$L$ & $D_c^*$ & $K$ & truncation errors\\
\colrule
32       & 0.96774204157 & 1500 & $1.1\times 10^{-10}$ \\
40       & 0.96800455535 & 1500 & $2.5\times 10^{-10}$ \\
48       & 0.96814719145 & 1500 & $3.7\times 10^{-10}$ \\
56       & 0.96823319962 & 2000 & $2.3\times 10^{-10}$ \\
64       & 0.96828901845 & 2000 & $4.8\times 10^{-10}$ \\
$\infty$ & 0.96847133(2)\footnote{linear fitting} && \\
$\infty$ & 0.96847141(2)\footnote{parabolic fitting} && 
\end{tabular}
\end{ruledtabular}
\end{table}

\subsection*{Level spectroscopy for the Haldane-Large-\textit{D} critical point}
The continuous quantum phase transitions are possible detected by the level crossing of excited states. The LS has been performed by different numerical algorithm with large system sizes, \textit{e.g.}, LS+DMRG~\cite{Tzeng2012} and LS+QMC~\cite{Suwa2015}. In the Table~\ref{tab:LS} , we show the detail numerical data for the Fig.~\ref{Fig:LS}. The errors in the Lanczos algorithm for the ground state energies are smaller than $10^{-13}$, and the dimension of the $(m,p,t)$-subspace for the Lanczos diagonalization is about $4\times 10^7$. Four sweeps are performed. A comparison of previous determination of $(D/J)_c$ is listed in Table~\ref{tab:Dc}.

\subsection*{Symmetry of parameters in Heisenberg models with the single-ion anisotropies}
In this section, we will discuss the symmetries of parameters in the Hamiltonian given by
\begin{equation}\label{eq1}
  H=J\sum_{\langle i,j\rangle}\vec S_i \cdot \vec S_j+D\sum_i(S_i^z)^2+E\sum_i[(S_i^x)^2-(S_i^y)^2].
\end{equation}
First, we can easily see that the model has a symmetry with respected to the sign flip of $E$ because the sign of $E$ can be absorbed into the exchange of $S^x$ and $S^y$. In addition to this symmetry, the Hamiltonian has a symmetry with respected to the permutation of spin operators. Note that the Heisenberg interaction is unchanging with respected to the permutation of $S^x$, $S^y$ and $S^z$ operators. Thus, the ground states of the model share qualitatively same properties among the models obtained by permuting the $S^x$, $S^y$, and $S^z$ terms in the single-ion anisotropy terms: such new models are
\begin{equation}
  H'=J\sum_{\langle i,j\rangle}\vec S_i \cdot \vec S_j+D\sum_i(S_i^x)^2+E\sum_i[(S_i^y)^2-(S_i^z)^2],
\end{equation}
and
\begin{equation}
  H''=J\sum_{\langle i,j\rangle}\vec S_i \cdot \vec S_j+D\sum_i(S_i^y)^2+E\sum_i[(S_i^z)^2-(S_i^x)^2].
\end{equation}
Because of this permutation symmetry, the ground-state phase diagram of the model has the symmetry in the parameter space spanned by $D$ and $E$. In the following, we investigate the symmetry in the phase diagram.

\begin{table}[t]
\caption{\label{tab:Dc}
A comparison of the $(D/J)_c$ determination.}
\begin{ruledtabular}
\begin{tabular}{l c c r}
$(D/J)_c$ & Method & Reference & Year\\
\colrule
0.97 & fidelity (DMRG) & \citet{Tzeng2008a} & 2008\\
0.971(5) & stiffness (QMC) & \citet{QMC} & 2009\\
0.96845(8) & entropy (DMRG) & \citet{accurate} & 2011\\
0.9684713(1) & LS+DMRG & this work & 2017
\end{tabular}
\end{ruledtabular}
\end{table}

In order to investigate the permutation symmetry, here we focus on the single-ion anisotropy terms and consider the one body Hamiltonian
\begin{equation}
  H_{\mathrm{SA}}=DS_z^2+E(S_x^2-S_y^2).
\end{equation}
By using the identity relation
\begin{equation}
  S^2=S_x^2+S_y^2+S_z^2,
\end{equation}
we can remove one of the spin operators ($S_x$, $S_y$ or $S_z$) from the Hamiltonian. Thus, we have three representations of the Hamiltonian as
\begin{align}
  H_{\mathrm{SA}}&=(E-D)S^2_x-(E+D)S^2_y+DS^2,\label{eq6}\\
  H_{\mathrm{SA}}&=-2ES^2_y-(E-D)S^2_z+ES^2,\label{eq7}
\end{align}
and
\begin{equation}\label{eq8}
  H_{\mathrm{SA}}=(D+E)S^2_z+2ES^2_x-ES^2.
\end{equation}
Note that the terms proportional to $S^2$ is constant and it does not change the ground state. Thus, hereafter, we neglect these constant terms for the simplicity.

From these three representations, we can easily see that the model with $E=0$ ($H_{\mathrm{SA}}=DS^2_z$) corresponds to the models with $E=D$ ($H_{\mathrm{SA}}=-2ES^2_y$) and $E=-D$ ($H_{\mathrm{SA}}=2ES^2_x$). Thus, the topological phase transition between the Haldane phase and the Large-$D$ phase at $D_c$ is mapped on two points $(D,E)=(\frac{1}{2}D_c,\pm\frac{1}{2}D_c)$.

In addition to this mapping, we can see a ``rotational'' symmetry by introducing a rescaled parameter $\tilde{E}=\sqrt{3}E$. In order to see the rotational symmetry, we introduce the polar coordinate in the parameter space $(D,\tilde{E})$ as
\begin{equation}
  (D,\tilde{E})=r(\cos\theta,\sin\theta).
\end{equation}
By substituting this relation into Eq.~(\ref{eq6}), we obtain
\begin{equation}\label{eq10}
  H_{\mathrm{SA}}=\frac{2r}{\sqrt{3}}\left[\sin(\theta-\frac{\pi}{3})S_x^2
  -\sin(\theta+\frac{\pi}{3})S_y^2\right].
\end{equation}
In the same way, we transform Eqs.~(\ref{eq7}) and (\ref{eq8}) as
\begin{equation}\label{eq11}
  H_{\mathrm{SA}}=\frac{2r}{\sqrt{3}}\left[ -\sin(\theta)S_y^2-\sin(\theta-\frac{\pi}{3})S_z^2\right],
\end{equation}
and
\begin{equation}\label{eq12}
  H_{\mathrm{SA}}=\frac{2r}{\sqrt{3}}\left[\sin(\theta+\frac{\pi}{3})S_z^2+\sin(\theta)S_x^2\right],
\end{equation}
respectively. Now, we can see the rotational symmetry easily. If we rotate the parameter by 120 degrees as $\theta\to\theta+\frac{2\pi}{3}$, Eq.~(\ref{eq10}) becomes
\begin{equation}
  H_{\mathrm{SA}}=\frac{2r}{\sqrt{3}}\left[\sin(\theta+\frac{\pi}{3})S_x^2+\sin(\theta)S_y^2\right].
\end{equation}
This is equivalent to Eq.~(\ref{eq12}) by permuting spin operators as $S_x\to S_z$, $S_y\to S_x$ and $S_z\to S_y$. In the same way, by $-120$ degrees rotation, $\theta\to\theta-\frac{2\pi}{3}$, we obtain
\begin{equation}
  H_{\mathrm{SA}}=\frac{2r}{\sqrt{3}}\left[-\sin(\theta)S_x^2-\sin(\theta-\frac{\pi}{3})S_y^2\right].
\end{equation}
This is equivalent to Eq.~(\ref{eq11}) by permuting spin operators as $S_x\to S_y$, $S_y\to S_z$ and $S_z\to S_x$.

As we emphasized previously, the Heisenberg coupling is unchanging under such permutations of spin operators. Thus, when we consider the phase diagram of the model Eq.~(\ref{eq1}) in $(D,\tilde{E})$ plane, the phase boundaries must have the symmetry with respected to 120 degrees rotation. We can also map the characteristics of each phase by considering the permutations of spin operators.

\end{document}